\documentclass{pasa}%

\usepackage{graphicx}
\usepackage{multirow,natbib}
\usepackage{textcomp}
\usepackage{epsfig}
\usepackage{amsmath}
\usepackage{amssymb}
\usepackage{amsthm}
\usepackage{hhline}
\usepackage{xy}
\usepackage{pbox}
\usepackage{xcolor}
\usepackage{mathtools}
\usepackage{longtable}

\usepackage{ulem}

\title{Search for Intermittent X-ray Pulsations from Neutron Stars in Low Mass X-ray Binaries}

\author[Bahar et al.]{Yunus Emre Bahar$^1$\thanks{baharyunus@sabanciuniv.edu}, Manoneeta Chakraborty$^2$ and , Ersin  G\"o\u{g}\"u\c{s}$^1$
\affil{$^1$Sabanc\i~ University, Faculty of Engineering and Natural Sciences, Orhanl\i ~Tuzla 34956 Istanbul Turkey}%
\affil{$^2$DAASE, Indian Institute of Technology Indore, Khandwa Road, Simrol Indore 453552, India}
}%

\jid{PASA}
\doi{10.1017/pas.\the\year.xxx}
\jyear{\the\year}

\usepackage{aas_macros}
\usepackage{hyperref} 
\hypersetup{colorlinks,citecolor=blue,linkcolor=blue,urlcolor=blue}

\hypersetup{draft}

\begin{document}

\begin{frontmatter}
\maketitle

\begin{abstract}
We present the results of our extensive binary orbital motion corrected pulsation search for 13 low mass X-ray binaries (LMXBs). These selected sources exhibit burst oscillations in X-rays with frequencies ranging from 45 to 1122 Hz, and have a binary orbital period varying from 2.1 to 18.9 hours. We first determined episodes that contain weak pulsations around the burst oscillation frequency by searching all archival Rossi X-ray Timing Explorer (RXTE) data of these sources. Then, we applied Doppler corrections to these pulsation episodes to discard the smearing effect of the binary orbital motion and searched for recovered pulsations at the second stage. Here we report 75 pulsation episodes that contain weak but coherent pulsations around the burst oscillation frequency. Furthermore, we report eight new episodes that show relatively strong pulsations in the binary orbital motion corrected data.
\end{abstract}

\begin{keywords}
methods: data analysis – stars: neutron – X-rays: binaries – X-rays: stars
\end{keywords}
\end{frontmatter}

\section{Introduction}\label{Introduction}

Neutron stars in low mass X-ray binaries (LMXBs) are among the brightest sources in the X-ray sky. Their emission is predominantly powered by the accretion of matter from the Roche lobe filled late type companion star; namely, the conversion of the gravitational potential energy of the infalling material into radiation near or on the surface of the compact object. As a result, such neutron stars either emit continuously (persistent sources) or only when the accretion mechanism is reinstated (transient sources). There are plethora of characteristic observational imprints in their X-ray emission, such as twin quasi-periodic oscillations (QPOs) in the kHz regime \citep{Klis1997}, thermonuclear X-ray bursts with oscillations \citep{Watts2012}. However, the majority of nearly 200 known neutron stars in LMXBs do not emit coherent X-ray pulsations.

Accreting millisecond X-ray pulsars (AMXPs) has emerged as a subclass of neutron star in LMXBs in the last two decades \citep{PatrunoWatts2012}. These transient systems exhibit coherent pulsations with periods shorter than ${\sim}10$ ms. The remarkable capability of AMXPs to turn the accretion energy into pulsations raises a question regarding the reason behind the lack of X-ray pulsations from the majority of the LMXBs. There might be no emerging pulsations from majority of LMXBs possibly because of insufficient magnetic field strengths to channel the accreting matter to the magnetic pole. Alternatively, the beamed signal could already be present but weak; it might be further weakened to non-detection while emerging pulsed radiation undergoes various dynamical or physical processes. Detection of intermittent episodes of coherent pulsations in the persistent emission phase from couple of systems (Aql X--1, \citet{Casella2008}; SAX J1748.9--2021, \citet{Altamirano2008}) support the case that all neutron stars in LMXBs are likely to radiate X-ray pulse. These observations provided a unique opportunity to better understand the reason of appearance and then disappearance of the pulsed emission.

 There are various neutron star atmosphere and surroundings effects that can reduce the pulse amplitude, and in turn, lead to the lack of pulsations. The most notable scenarios for this situation include light bending effect resulting from the extreme gravity of the compact object \citep{Wood1988, Ozel2009}, scattering characteristics of the environment surrounding the neutron star \citep{bra87, tit02}, and the magnetohydrodynamic instabilities in the accretion flow \citep{kul08}. Another possible cause is the binary orbital motion, which could smear the already weak signal to non-detection. Such a dynamical effect would introduce significant implications on the pulsed signal especially in tight binary systems. In such systems, it could, in principle, be possible to recover a pulsed signal in the X-ray data if the binary orbital effects are accounted for.

There have been several extensive studies to correct smearing of pulsations due to dynamical effects from binary neutron stars. One approach, the acceleration search provides a partial correction by splitting the data into short segments and assuming that the orbital acceleration of the neutron star to be approximately constant during the segments of the binary orbit which corresponds to the exposure time of the obsevation \citep[e.g.][]{Middleditch1984,Anderson1990,Wood1991}.
Later, \citet{Ransom2001,Ransom2002} elaborated this technique to obtain template responses in the frequency - frequency derivative ($f$-$\dot{f}$) plane. They correlate the Fourier amplitude and phase responses of the real time data with template responses that are generated with trial acceleration values. Recently, it was extended by assuming jerk ($\ddot{f}$) to be constant, instead of the orbital acceleration \citep{Andersen2018}. By doing that they added one more dimension to their parameter space and conducted their search in $f$-$\dot{f}$-$\ddot{f}$ volume which they called jerk search \citep{Andersen2018}. These methods are widely employed to search for periodic signal in timing data collected in the radio band.

Another approach to detect smeared pulsations is called semi-coherent search. It was first proposed as a technique to detect weak gravitational wave signals \citep{Messenger2011}, and then applied to search for weak X-ray signals from neutron stars in LMXBs \citep{Messenger2015, Patruno2018}. The focus of this method is to detect weak but continuous pulsations. They attempted to achieve this goal by applying a two-step procedure on the X-ray data. In the first stage, they divide data to segments and process each segment with a bank of templates coherently. These templates are produced such that they account for the Doppler modulation in the phase of the signal by a Taylor expansion in frequency. In the second stage they incoherently combine the coherent signal power results obtained for each segment. They applied this technique to 12 LMXBs and no evidence was found for a previously non-detected weak pulsation.

Physical mechanism behind pulse smearing is the relativistic Doppler effects caused by the rotational motion of the neutron star around the common center of mass of the binary system. Effectively, Doppler effects modulate the signal by causing delays on the photon arrival times depending on the relative position of the neutron star on its orbit. This delay was formulated by \citet{Blandford1976} for observing and testing various relativistic effects by making use of the information that pulsars are reliable and precise clocks. In this study, we use this formulation to revert the effect of the binary orbital motion to possibly strengthen the pulsed signal. Smearing can redistribute the signal power to other frequencies around the actual frequency such that the power of the smeared signal lies near or below the detection threshold. We, therefore, applied a systematic investigation by first correcting the arrival times of photons including pulsed signals that are just below the detection level, and then searched for periodic signal in the corrected data. Note that earlier strategies to search for weak pulsations relied on sophisticated templates to model the binary effects on the coherent pulsation signal. In our method, we first account for the smearing effect by transforming the photon arrival times to an inertial frame at the common center of mass of the binary system and then use a simple sinusoidal model to measure the strength and the frequency of the pulsed signal.

Here we present the results of our extensive search for transient episodes of pulsed X-ray emission in the entire Rossi X-ray Timing Explorer (RXTE) data of 13 binary neutron stars systems, as well as our investigations to constrain the orbital characteristics of these systems.

\begin{table*}
    \centering
    \caption{\label{fpt_table} Fundamental characteristics and RXTE observational details of the systems investigated}
    \begin{tabular}{lccccc}\hline
    Source  & Burst Oscillation    & Orbital & Average Count Rate & Total Time & References \\ 
& Frequency ${f_{s}}$ (Hz)    & Period ${P_{b}}$ (Hr) & (${counts \ s^{-1}} \ PCU^{-1} $)\textsuperscript{b}  & Searched ${t_{tot}}$ (ks) & \\\hline
EXO 0748--676 &  45/552\textsuperscript{a}     &  3.82  & 38.4    & 2228.5   &  1, 2         \\
IGR J17191--2821 &  294     &  ... & 113.4 &  82.9 &  3                     \\
4U 1702-429 &  329          &  ... &  141.3   &  1250.4   &  4             \\
4U 1728-34 &  363           &  ... &  287.8   &  1649.3   &  5             \\
KS 1731-260 &  524          &  ... & 155.0 &  468.9 &  6, 7                 \\
A 1744-361 &  530           &  ... & 140.2 &  117.7 &  8                    \\
Aql X--1 &  550             &  18.95 &   368.0  &  1646.6   &  9, 10, 11, 12    \\
MXB 1658-298 &  567         &  7.11 & 73.9 &  339.0   &  13               \\
4U 1636-536 &  581          &  3.80  &   248.0  &  4384.7   &  14, 15          \\
SAX J1750.8-2900 &  601     &  ... &   179.6  &  214.7   &  16, 17         \\
GS 1826-238 &  611          &  2.10 &  124.6 &  1012.4   &  18              \\
4U 1608-52 &  620           &  12.89 & 349.1 & 2113.2    &  17, 19, 20           \\
XTE 1739-285 &  1122        &  ... &  91.5   &  118.9   &  21            \\        
\hline
\end{tabular}\\
\begin{flushleft} \footnotesize 
$^a$ There are two different burst oscillation frequencies reported for EXO 0748--676. \\
$^b$ Average count rates are calculated in the energy range of $\sim$3-27 keV from the all available X-ray data excluding thermonuclear bursts, that is, segment from 20 s before till 200 s after the peak of X-ray bursts. \\
\vspace*{0.1cm}
\textbf{References--} 1. \cite{Villarreal2004};
2. \cite{Galloway2010};
3. \cite{Altamirano2010};
4. \cite{Markwardt1999}; 5. \cite{Strohmayer1996}; 6. \cite{Smith1997};
7. \cite{Muno2000}; 8. \cite{Bhattacharyya2006}; 
9. \cite{Chevalier1991};
10. \cite{Zhang1998}; 11. \cite{Welsh2000}; 12. \cite{Casella2008};
13. \cite{Wijnands2001};
14. \cite{Strohmayer1998};
15. \cite{Strohmayer2002};
16. \cite{Kaaret2002};
17. \cite{Galloway2008};
18. \cite{Thompson2005};
19. \cite{Hartman2003};
20. \cite{Wachter2002};
21. \cite{Kaaret2007};
\end{flushleft}
\end{table*}

\section{Methodology}

We performed the search for weak pulsations in two steps. In the first tier, we conducted pulse search on the barycentered data and determined the candidates. In the second step, we applied arrival time corrections to these candidates to account for the effects of the binary motion and re-searched for pulsations in the corrected data.

\subsection{Observations, the $1\textsuperscript{st}$ Tier Search and Results}\label{presearch}
For the first tier pulsation search, we selected 13 neutron star low mass X-ray binary and used all available event mode archival RXTE data. Note that none of these sources show X-ray pulsations above the noise level during their persistent emission phase (except Aql X--1). However, ten of them show prominent quasi-periodic oscillations just before, during or just after they ignite a thermonuclear X-ray burst. Remaining three also have reported periodicities in their burst observations. Nevertheless, these sources (XTE 1739--285, A 1744--361, GS 1826--238) show either only one burst or the burst oscillations are tentative. We assume that thermonuclear burst oscillations correspond to an X-ray emitting hotspot and the spin frequency of the underlying source is around the frequency of these oscillations (see \citet{Watts2012}). Binary orbital periods of six of these sources are known either from periodicities or eclipses observed in the X-ray data. We present the list of the sources investigated and their burst oscillation frequencies in Table~\ref{fpt_table}. 

Before applying the first tier search, we generated the light curve of each RXTE pointing in the 2--60 keV energy band with 0.125 s time resolution to search for thermonuclear X-ray bursts. We then created good time intervals for each source by excluding the times of identified X-ray bursts. In particular, we excluded the data starting 20 seconds before the burst peak till 200 seconds after. This conservative selection excludes any possible contribution from even the relatively longer duration bursts. Note that the source 4U 1728--34 has a type-II X-ray burster (the Rapid Burster) in its RXTE PCA field of view.
For this system, we ignored all observations with type-II bursts present and excluded them from our list of good time intervals. We list the total investigated observing time for each source after the exclusion of the burst intervals in Table~\ref{fpt_table}. Finally, we transferred the photon arrival times to the Solar System barycenter to get rid of the relativistic effects of the moving frame of the detector. 

In the first tier search, we fixed channel ranges from one observation to the other rather than fixing energy ranges because energy-channel relation of RXTE has changed during its lifetime. This did not create any problem since we have not compared any observation to the other directly. We applied statistical analysis techniques for comparing any result to the other.

To make our search sensitive for the pulsations that are made up of only hard or low energy X-ray photons, we carried out the first step of our pulsation search in three energy bands; $\sim$3-9 keV (absolute channels of 7-24), $\sim$9-27 keV (channel range of 25-70), and $\sim$3-27 keV (7-70 channels). Here we note that the energy ranges may vary slightly between the various gain epochs as we have kept the channel range fixed. For each energy band, we constructed a 256 second window at the very beginning of each observation and generated a lightcurve from the photons that are within this window with a 1/2048 s binning. This bin size corresponds to a maximum frequency of 1024 Hz in the Fourier domain and it is above all of the reported LMXB burst oscillation frequencies except XTE 1739--285 for which we used 1/4096 s binning. We constructed the Leahy normalized power density spectrum \citep{Leahy1983} from this lightcurve and calculated the statistical significance of the maximum power between $f_{s}\pm 10$ Hz where $f_{s}$ is the reported burst oscillation frequency. Note the fact that intermittent weak pulsations are non-stationary by definition. This makes it harder to decide how to deal with the number of trials while converting the signal powers to its corresponding statistical significance values. We decided to calculate statistical significance values by considering each power spectrum on its own and avoid confusion by providing the power levels besides the significance levels.
Therefore, first the single trial probability of obtaining the highest Leahy power is found and then joint probability of having the spectrum is calculated with the number of trials ($N_{trials}$) equal to the number of frequency bins searched i.e., the number of frequency bins between $f_{s}\pm 10$ Hz. At last, significance level of this joint probability is calculated from a normal distribution in the light of the central limit theorem. Here we have calculated the Gaussian significance of the joint probability by considering a two-tailed test given the p-value. It is to be noted that the significance values do not change by an appreciable amount if one-sided test (considering one tail of the Gaussian) is applied. We then slided the 256 s interval by 16 seconds, repeated the same procedure to obtain the significance for that time interval, and continued until the end of the observation. This procedure would facilitate the detection and strengthening of any signal which is present of a short time duration.

After calculating the statistical significance of the strongest pulse for each window, we selected candidates by applying the following continuity, coherence and minimum significance criterion. We require a pulsation candidate for further detailed analysis to have at least 2.5${\sigma}$ statistical significance for four consecutive time segments and having the maximum Leahy power between $f_{s}\pm 2$ Hz. By setting this criterion we aimed to uncover coherent periodicities around the burst oscillation frequency that are just below the detection threshold. Our first tier search resulted in 75 episodes of pulsation candidates from 10 sources. We also found that the search in the broader energy range resulted in the same pulsation candidates in the lower and upper energy bands. We, therefore, continued our investigations within the absolute channel range of 7 to 70 ($\sim$3 - 27 keV).

\begin{figure*}
\centering
\begin{tabular}{c}
\includegraphics[width=\textwidth]{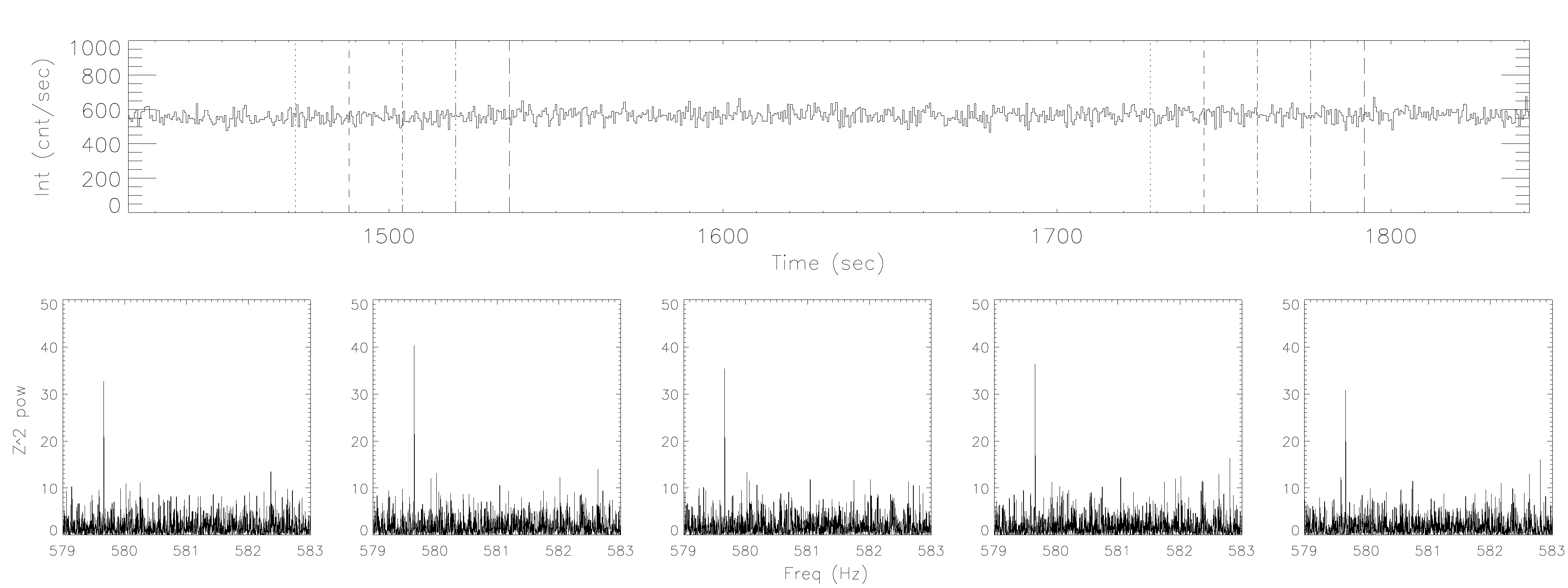}
\end{tabular}
\caption{Example of a candidate pulsation episode from 4U 1636--536 that lasts for five consecutive time windows. The pulsation candidate and its properties are indicated in boldface in Table~\ref{prelim_res_table}. {\it (top panel)} The light curve of the part of the observation that contains 336 s long candidate pulsation episode. The vertical dashed lines that have the same line style correspond to the starting and ending time of 256 s windows from which power spectra are calculated. {\it (bottom plots)} ${Z^2}$ power density spectra of the five sequential 256 s intervals indicated with vertical lines above. The signal at 579.66 Hz is clearly evident in all plots.
\label{fig X13}}
\end{figure*}

Conventional fast Fourier transform (FFT) can only be applied to discrete data with equally spaced time axis. For that reason, photon arrival times are binned and histograms are created to apply FFT. Even though this approach is computationally very effective, it also has downsides. To be able to bin the data one need to choose the starting time of the binning according to the first photon of interest. Power spectrum may slightly change depending on the choice of the starting time of the binning because shifting the starting and ending times of the histogram bins can cause the photons to be redistributed which will change the values of the histogram. For these reasons, we confirmed the pulsation candidates by using a $Z^2$ test \citep{Buccheri1983} which is computationally more expensive but it does not require the data to be binned. The $Z^2$ power is formulated as

\begin{equation}
Z^2_n = \frac{2}{N} \sum_{m=1}^{n} [\{\sum_{j=1}^{N} \cos (m\phi_j)\}^2 +\{\sum_{j=1}^{N} \sin (m\phi_j)\}^2 ]
\end{equation} 
where $n$ is the number of harmonics, $N$ is the total number of photons, and $\phi_j$ is the phase of the $j^{\rm th}$ photon. $Z^2$ powers are distributed with a ${{\chi}^2}$ distribution with $2n$ degrees of freedom where $n$ corresponds to the desired number of harmonics. In our case, we have chosen the number of harmonics to be 1 since strong harmonics are not reported for the sources of interest. The $Z^2$ powers are calculated with a 1/512 Hz frequency resolution between $f_{s}\pm 2$ Hz. Significance values are calculated with the same approach as above. We present the results of our first tier search and the corresponding pulsation candidates in Table~\ref{prelim_res_table}. We present an example case for a pulsation episode in Figure~\ref{fig X13}. As in this case, none of our candidate pulsation episodes contain burst emission.

\subsection{Binary Orbital Motion Corrected Search} \label{Detailedsearch}

Until this point, we searched for coherent pulsations in the data that only the barycentric correction was applied. However, binary orbital motion of the neutron star could smear out an already weak signal and make it undetectable. In our study, we used the appropriate relativistic orbit model \citep{Blandford1976} for the Doppler correction to be able to recover the smeared signal based on plausible orbital parameters. The time delay, ${t_{d}}$ due to orbital motion is formulated as

\begin{equation} \label{eq:Kirchhoff's 3}
\begin{split}
 t_{d} &=  \left \{ \alpha \left ( \cos E - e \right ) + \left (\beta + \gamma   \right)\sin E \right \} \\ & \qquad \qquad \qquad \ \  \times \left \{ 1 - \frac{2 \pi}{P_{b}}  \frac{ \beta \cos E - \alpha \sin E }{ 1- e \cos E } \right \}
\end{split}
\end{equation}
where $\alpha = x \sin w$ and $\beta =  \left ( 1 - e^{2} \right )^{1/2} x \cos w$ are expressed in terms of the projected semi-major axis, $x = a \sin i$, ${P_{b}}$ is the binary period, ${E}$ is the eccentric anomaly, ${e}$ is the eccentricity, ${w}$ is the longitude of periastron and ${\gamma}$ is the term for the gravitational redshift and time dilation. The eccentric anomaly ${E}$ is defined as

\begin{equation}
E - e\sin E = \frac{2\pi t}{P_{b}}
\end{equation} 
from the Kepler's equation \citep{Taylor1989}. In our search we assumed the orbit of the neutron star to be circular which is a plausible approximation for LMXBs ($e = 0$).
This assumption reduced the number of free parameters to three that are the binary orbital period (${P_{b}}$), projected semi-major axis (${x}$) and the epoch of mean longitude equal to zero ($T_{0}$). Because of the lack of information and uncertainties about the epoch of mean longitude equal to zero, we decided to apply a ${\pi/4}$ radian sampling to the phase (${\psi}$) of the circular orbit that corresponds to 8 trial phases in total. 
This way, we aimed to test roughly all possible configurations that can effect the signal. For the six sources whose binary orbital periods were already reported (see Table~\ref{fpt_table}), we limited the search interval of binary period to within 1 hr around their reported value, with a 0.1 hr sampling in those 2 hr intervals. For the rest of the sources whose binary period is unknown, we chose the range of the trial range of orbital periods to be 2--30.5 hr and applied 1.5 hr sampling to this time interval. This orbital period interval is expected to correspond to a great majority of LMXB orbital periods \citep{vanHaaften15}. Similar to the uncertainties about the epoch of mean longitude equal to zero ($T_{0}$) of the sources, projected semi-major axis (${x}$) values of LMXB sources were also either unknown or highly uncertain. For this reason, we select a broad range of trial values from 0.01 to 1.91 light-s with 0.1 light-s sampling. Therefore, for each 256 s pulsation candidate time segment, the correction is applied for $8\times20\times20$ sets of parameters which span ${T_{0}}$, ${P_{b}}$ and ${a\sin i}$ parameters, respectively. Sampling sizes of the correction parameters were limited mainly due to the computational costs. We tried to cover a wide range of parameters with the smallest grid size possible given the present computational capabilities. We would like to also note that the reported binary periods of our six sources are much greater than 256 s. In practice, one could reduce the parameter space further based on this fact (see \citet{Ransom2001,Ransom2002}). However, we decided to pursue with the \citet{Blandford1976} correction and applied it in all three dimensions for the sake of completeness and for creating a generalized approach applicable to different sources.

We then searched for pulsations in the binary orbital motion corrected data by employing the Z$^2$ method between f$_{s} \pm$ 2 Hz as done before. By doing that, every correction that is applied with a different parameter set gave different statistical significance values for each time segment. Afterwards, we calculated the highest frequency shift that can occur (${\Delta f_{max}}$) from the first order calculations

\begin{equation}\label{mfreqshift}
\Delta f_{max} = f_{p} \left | \frac{P_{b}}{{P_{b}}- 2\pi a \sin i} - 1 \right|
\end{equation} 

and then located recovered pulsations by applying the following set of criterion: We require the recovered pulsations to have at least 3.5, 4.5 and 3.5${\sigma}$ statistical significance for three consecutive time segments and we require significance values for these three segments to be higher than before (that is, before the binary motion correction). We choose these significance levels since we wanted to put a significance criterion at least 1${\sigma}$ higher than the previous (2.5${\sigma}$) for the detection of the recovered pulses. We require the improvement in the significance levels for three time segments to be achieved concurrently after corrected with the same parameter set. We also require the pulsations to be at the same frequency for these three time segments and the frequency of the recovered pulsation to be at most ${\Delta f_{max}}$ away from the frequency of the pulsation before correction.

After applying this set of criterion we identified recovered pulsations for five sources; EXO 0748--676, 4U 1608--52, 4U 1636--536, Aql X--1, 4U 1728--34 out of 10 that showed pulsation candidates. We obtained degenerate pulsations for almost all time segments that show recovered pulsations. These pulsations were degenerate such that, for a single time segment there were many recovered pulsations that were obtained with different ${P_{b}}$, ${a\sin i}$, and ${\psi}$ parameter sets. Moreover, pulsation strengths of recovered pulsations for a single time segment were at a similar level. The presence of degenerate recovered pulsations was expected because of the degenerate nature of Doppler effects. For example, to obtain similar Doppler effects for a trial orbital phase there should be a relation between ${P_{b}}$ and ${a\sin i}$ such that if ${P_{b}}$ is short, ${a\sin i}$ should also be shorter to compensate each other.

Binary orbital period (${P_{b}}$) of four of the five sources were already known (except 4U 1728--34). This helped us to eliminate the degenerate cases of recovered pulsations. We discarded the recovered pulsations that are obtained with a ${P_{b}}$ that is different than the reported ${P_{b}}$. Furthermore, for each time segment we selected the parameter set that gives the the highest signal power among the ones that have the appropriate ${P_{b}}$. Since we did not know the binary period of 4U 1728--34, we adopted a slightly different approach. Since sampling for the ${P_{b}}$ parameter of 4U 1728--34 is 15 times larger than the ones that the binary periods are known, we decided to decrease the significance criterion of the detection of the recovered pulsation to 3.5, 4.0 and 3.5 from 3.5, 4.5 and 3.5. By doing so, we aimed to report more recovered pulsations and corresponding ${P_{b}}$, ${a\sin i}$ couples for such a highly unknown situation. We list the results of the recovered pulsations and the corresponding correction parameters in Table~\ref{binary_motion_corrected_search_res_table}. Note that we also report in Table~\ref{binary_motion_corrected_search_res_table} the binary motion corrected results for the previously detected 150 s intermittent pulsation episode from Aql X--1 \citep{Casella2008}. Each line in this table corresponds to a 256 second time window that meets our detection criteria and it can be seen that some of these time windows are adjacent to each other. We count the adjacent windows as one pulse episode which makes a total of eight new pulsation episodes reported in the table excluding the detection of \citet{Casella2008}.

\begin{table*}
    \centering
    \caption{\label{binary_motion_corrected_search_res_table} Results of the binary motion corrected search}
    \begin{tabular}{llccccc}\hline
    Source  & Time (UTC)\textsuperscript{a} & ${P_{b}}$ & ${a\sin i}$& ${f_{Z^{2}}}$  & ${Z^2}$ & ${sig_{Z^{2}}}$    \\
 &  & (Hr) & (lt-s) & (Hz) & Power& (${\sigma}$) \\\hline
EXO 0748--676 &                            1998 Mar 14 01:02:20.9 &      3.82 &      1.81 &       44.48 &       41.30 &        4.73  \\
 &                            2009 Jul 28 08:30:34.0 &      3.82 &      1.71 &       46.43 &       43.72 &        4.97  \\
 \hline
4U 1608--52 &                             2007 Nov 1 06:39:58.1 &     12.89 &      1.91 &      619.28 &       39.42 &        4.54  \\
\hline
4U 1636--536 &                             2002 Jan 8 18:13:46.4 &      3.80 &      0.61 &      580.57 &       40.33 &        4.63  \\
 &                            2006 Apr 23 11:21:54.8 &      3.80 &      0.61 &      579.82 &       41.17 &        4.72  \\
 &                            2006 Apr 23 11:22:26.8 &      3.80 &      1.71 &      580.12 &       40.05 &        4.61  \\
 \hline
Aql X--1\textsuperscript{b} &                            1998 Mar 10 22:48:34.9 &     18.95 &      1.71 &      550.22 &       56.71 &        6.11  \\
 &                            1998 Mar 10 22:48:50.9 &     18.95 &      1.71 &      550.22 &       68.10 &        6.96  \\
 &                            1998 Mar 10 22:49:06.9 &     18.95 &      1.91 &      550.27 &       77.36 &        7.59  \\
 &                            1998 Mar 10 22:49:22.9 &     18.95 &      1.91 &      550.28 &       88.84 &        8.30  \\
 &                            1998 Mar 10 22:49:38.9 &     18.95 &      1.11 &      550.27 &       97.97 &        8.83  \\
 \hhline{=======}
 4U 1728--34\textsuperscript{c} &                            1997 Sep 24 12:59:41.1 &      3.50 &      0.91 &      364.42 &       35.84 &        4.15  \\
  &                            1997 Sep 24 12:59:57.1 &      2.00 &      0.31 &      364.23 &       35.04 &        4.05  \\
  &                            1997 Sep 24 13:00:13.1 &      5.00 &      1.31 &      364.19 &       36.14 &        4.18  \\
  &                            1997 Sep 24 13:00:29.1 &      2.00 &      0.91 &      364.60 &       35.55 &        4.11  \\
  &                            1999 Feb 28 02:41:15.9 &      8.00 &      1.81 &      363.37 &       39.40 &        4.54  \\
  &                             2002 Mar 5 12:40:52.8 &      6.50 &      1.11 &      361.94 &       36.93 &        4.27  \\
  &                             2002 Mar 5 12:41:08.8 &      3.50 &      1.11 &      362.06 &       35.17 &        4.07  \\
  &                             2002 Mar 5 12:41:24.8 &      3.50 &      1.61 &      362.15 &       36.05 &        4.17 \\
    
\hline
\end{tabular}\\
\begin{flushleft} \footnotesize 
$^a$ Time (UTC) is the starting time of the 256 s time window. \\
$^b$ Lines under Aql X--1 are orbital motion corrected results of the previously discovered 150 s pulsation episode by \citet{Casella2008}\\
$^c$ Detection criterion of recovered pulsations for 4U 1728--34 is different than other sources which is 3.5, 4.0 and 3.5 rather than 3.5, 4.5 and 3.5 for three consecutive time segments. See Section~\ref{Detailedsearch} for a more detailed discussion. \\
\end{flushleft}
\end{table*}

There were two different spin frequencies in literature that thought to be related with the spin frequency of EXO 0748--676 that are 45 and 552 Hz. We were able to obtain pulsation candidates at both of these frequencies in our first tier search. However, after applying and correcting the photon arrival times, we obtained recovered pulsations only around 45 Hz even though smearing effect of the Doppler modulation is increasing at higher frequencies. 

\section{Discussion}

We have performed one of the most comprehensive search for intermittent pulsation episodes in non-pulsing LMXBs. In particular, we have accounted for the binary orbital motion systematically in our extensive investigations. Recently, \citet{Messenger2015} and \citet{Patruno2018} also accounted for the Doppler shift in searching pulsations from LMXBs with their semi-coherent search. Their work differ from ours at a fundamental level: They aimed to detect continuous but weak pulsations, contrary to our search for intermittent pulsation episodes. Furthermore, eight out of 12 of their LMXB sample did not show burst oscillations. This made their parameter space significantly wider for these  LMXBs contrary to more confined spin frequency parameter in our work thanks to the previously reported burst oscillations for the LMXBs. Note that we have nine different LMXBs in addition to four common with their work, namely; Aql X--1, 4U 1636--536, 4U 1608--52 and XTE 1739--285.

Detection of a single $\sim$150 s long pulsation episode \citep{Casella2008} in the entire archival RXTE data of Aql X--1's persistent emission phase clearly show that the intermittent pulsations are very rare (0.009\% of the total 1645 ks observed time). Note that the pulsation episodes in other sources are also rare with respect to the total observation time: 0.023\% for EXO 0748--676, 0.012\% for 4U 1608--52, 0.016\% for 4U 1636--536 and 0.062\% for 4U 1728--34. We should note here that occurrence rate for 4U 1728--34 is expected to be higher since we lowered significance criterion for this source. (See Section~\ref{Detailedsearch}).
Besides these recovered pulsations, we also found pulsation candidate episodes that contain weak pulsations around the burst oscillation frequencies for all 13 LMXBs in our sample. These candidate episodes are also short and rare compared to the total searched time of each source: 0.24\% for EXO 0748--676, 0.08\% for 4U 1608--52, 0.13\% for 4U 1636--536, 1.84\% for MXB 1658--298, 0.07\% for 4U 1702--429, 0.21\% for 4U 1728--34, 0.52\% for XTE 1739--285, 0.15\% for SAX J1750.8--2900, 0.12\% for GS 1826--238, and 0.19\% for Aql X--1

\cite{Strohmayer2018} employed the same binary orbital motion correction to account for Doppler smearing in an accreting millisecond X-ray pulsar system, IGR J1762--6143 with an ultra-compact orbit (binary orbital period is ${\sim}38$ min). They found that the orbital delays weaken the pulsation such that after the correction to its NICER data, $Z^2$ power of the pulsation increase up to ${\sim}196$ which is almost 4 times the previous value (${\sim}55$). However, their work differs from ours on a fundamental aspect. IGR J1762--6143 shows strong pulsations as side bands even when no correction was applied. Correcting the photons put the side band structure together and result with a stronger centered pulse. In our approach, we do not observe clear pulsations as side band structures in the non-corrected data since the strength of the pulsations that we have been looking for were not strong enough to be detected after smearing. Nevertheless, it is important to remark that side band structured pulses of IGR J1762--6143 clearly show that smearing effect of the binary orbital motion plays a key role in disappearance of already weak pulsations. This makes it almost mandatory to apply the Doppler correction to the photon arrival times before searching weak pulsations especially for compact (short ${P_{b}}$) sources considering the high uncertainties in the inclination angles.

We were able to determine orbital inclination of four systems, based upon our criterion that the search in the Doppler corrected data would yield the strongest signal. We suggest the projected semi-major axis (${a\sin i}$) value of about (5.1--5.4)$\times10^{8}$ m for EXO 0748--676, and $5.7\times10^{8}$ m for 4U 1608--52. For typical orbital separations ($a$) that can be calculated from the known binary periods, the orbital inclination of these two systems turned out as 29.0--30.9 and 13.9 degrees, respectively. Our search for projected semi-major axis of 4U 1636--536 and Aql X--1 concentrate around two values: $1.8\times10^{8}$ or $5.1\times10^{8}$ m for the former, and $3.3\times10^{8}$ or (5.1--5.7)$\times10^{8}$ for the latter. These correspond to 9.9 or 29.1 degrees of inclination for 4U 1636--536 and 6.2 or 9.6--10.7 degrees for Aql X--1. We would like to note that, donor mass and neutron star mass are assumed to be 0.5 and 1.4 $M_{\odot}$ respectively for calculating the orbital separations. Highly degenerate recovered pulsations and their scattered orbital parameters in the ${P_{b}}$ -- ${a\sin i}$ plane made it impossible to determine orbital parameter estimates for 4U 1728--34. Two recovered episodes of pulses for EXO 0748--676 were found at the frequencies of 44.5 and 46.4 Hz. We found pulsation candidate episodes around 552 Hz, which is the other suggested frequency, and also there are other studies supporting that \citep{Balman2009,Jain2011}. However, we did not detect any recovered pulsations after the Doppler correction around the higher frequency. In the light of these results, we argue that the 45 Hz may be the actual spin frequency of EXO 0748--676 rather than 552 Hz.

We have also investigated whether X-ray intensity (that is a measure of mass accretion rate) is different during those pulsation episodes. We compare the average count rate for each source (listed in Table 1) to the source rates during the episodes of coherent pulsations (listed in Table 2). We find no systematic trend in X-ray intensities, neither higher nor lower than their long term averaged values. We, therefore, suggest that the appearance of  pulsations are not linked to any sudden change in the accretion rate.

The lack of pulsations in these bright LMXB sources may not be solely related to the Doppler effects. There are already numerous suggestions attempting to explain the reason of intermittent pulsations. Sporadic behaviour and uncommon nature of such periodicities force the explanations to be somehow related with a rare incident or asymmetry that might be taking place close to the neutron star surface. These explanations can be roughly divided into two categories: one group of explanations assume that the pulsation is temporary while other group assume that the pulsation is always present however so weak that we cannot detect it.
 
One possible explanation for the lack of pulsations that belongs to the first group is the scattering of the beamed emission by the optically thick media. However, this approach was challenged by \citet{Gogus2007} where they argued through spectral investigations that the optical depth ${\tau}$ of the surrounding corona is not thick enough to smear out the pulsations. In order for this explanation to work, temperatures of scattering electrons should be very low ($\lesssim$10 keV) so that the optical depth would be large enough to screen. The pulsations would then become visible only during a spectral variability, in presence of a local hole in the screening medium or in presence of other unique asymmetric geometries \citep{Casella2008}. In such a scenario, one would also expect a dependence of the presence of the pulsation on the accretion rate as in such systems the coronal properties are strong function of the accretion rate. Such a correlation was found to absent in our analysis.
 
The second group of explanations is generally related to the magnetic channeling. Neutron stars in LMXBs are weakly magnetized. Therefore, they are typically unable to stop and focus the incoming matter. The appearance of pulsations could either be due to a sudden change in the amount of channeled matter or the strength of their magnetic field could suddenly increase (at least locally) and become capable of channeling matter. The second scenario might occur if there is a sudden decrease in the Ohmic diffusion time which could happen due to a starquake, local disruption of screening currents or magnetic reconnections \citep{Casella2008}. 
It is essential to study multipolar magnetic fields in such neutron stars to find an evidence for strengthening the magnetic fields. Recent serious efforts to understand magnetic field structure of neutron stars by ray-tracing technique \citep{Bilous2019} may provide new insights on this front.

Another approach within the second group explanations is nuclear burning. We expect nuclear powered oscillations to last typically a few seconds. However, \citet{Strohmayer2002} reported a long-lasting pulsation during a superburst of 4U 1636--536. This indicates that coherent and longer lasting oscillations may also be related with nuclear burning. This explanation cannot be related with our results since we carefully identified the bursts and eliminated their times from our searched sample. There could, in principle, be nuclear burning events which might not be seen radiatively (as thermonuclear bursts) but the stored energy could instead give rise to intermittent coherent oscillations.

\begin{acknowledgements}

We are grateful to the anonymous reviewer for conservative comments. We acknowledge support from the Scientific and Technological Research Council of Turkey (T\"UB\.ITAK, grant no: 115R034). 
\end{acknowledgements}

\onecolumn
\begin{longtable}{llcccccc}
\caption{Results of the first tier pulsation search}
\label{prelim_res_table}\\
\hline
\hline
    Source &  Time (UTC)\textsuperscript{a}  & ${t_{dur}}$ & Leahy   & ${Z^2}$ & ${f_{L}}$ & ${sig_{L}}$  & Count Rate   \\
    &  & (s)\textsuperscript{b} & Power\textsuperscript{c} & Power\textsuperscript{c} & (Hz)\textsuperscript{c} & (${\sigma}$)\textsuperscript{c}  & (${counts \ s^{-1}} \ PCU^{-1}$)\textsuperscript{d} \\\hline
    
    EXO 0748--676 &                        1998 Mar 14 01:02:04.9 &      320 &      35.9 &      35.7 &      44.51 &       3.94 & 36.8\\ 
 &                        1998 Jun 28 13:37:49.0 &      352 &      33.9 &      32.9 &      45.77 &       3.70 & 35.3 \\ 
 &                        2000 Mar 28 15:43:36.9 &      320 &      36.8 &      36.2 &      46.09 &       4.04 & 39.9 \\ 
 &                         2002 Sep 1 12:38:55.0 &      320 &      29.0 &      29.1 &      44.18 &       3.01 & 22.9 \\ 
 &                        2003 Feb 15 07:00:09.0 &      304 &      35.1 &      34.9 &      45.04 &       3.84 & 31.0 \\ 
 &                        2003 Aug 18 19:46:36.0 &      304 &      31.9 &      31.1 &      44.75 &       3.43 & 40.1 \\ 
 &                        2004 Apr 26 15:06:20.0 &      320 &      28.7 &      29.6 &      46.55 &       2.96 & 51.5 \\ 
 &                        2004 Apr 26 15:24:12.0 &      320 &      32.4 &      32.0 &      45.55 &       3.50 & 55.1 \\ 
 &                        2004 Nov 25 15:58:35.9 &      304 &      29.5 &      29.4 &      43.55 &       3.09 & 45.9 \\ 
 &                         2007 Feb 7 08:26:53.0 &      304 &      35.3 &      35.9 &      45.46 &       3.87 & 16.8 \\ 
 &                        2007 Aug 23 20:57:00.1 &      304 &      28.8 &      29.5 &      46.29 &       2.98 & 39.2 \\ 
 &                        2009 Jul 28 08:30:02.0 &      320 &      39.3 &      40.3 &      46.41 &       4.33 & 20.4 \\ 
 &                        1997 Jan 19 12:28:17.0 &      304 &      27.2 &      14.6 &     552.40 &       2.73 & 31.0 \\ 
 &                        2004 Sep 25 15:02:52.0 &      304 &      38.5 &      40.1 &     550.20 &       4.24 & 15.0 \\ 
 &                        2006 Sep 16 16:49:15.0 &      352 &      34.2 &      28.0 &     553.27 &       3.73 & 33.9 \\ 
 &                         2008 Feb 3 09:51:06.9 &      336 &      28.6 &      23.6 &     551.63 &       2.95 & 14.9 \\ 
 &                         2008 Feb 6 19:22:03.0 &      304 &      35.0 &      33.3 &     553.14 &       3.83 & 42.2 \\ 
\hline
4U 1608--52 &                         2002 Sep 1 09:25:30.0 &      304 &      32.0 &      28.6 &     620.13 &       3.45 & 746.3 \\ 
 &                        2002 Sep 28 18:26:26.0 &      320 &      31.7 &      29.7 &     620.53 &       3.40 & 46.2 \\ 
 &                         2003 Oct 4 17:11:39.0 &      304 &      30.6 &      20.6 &     620.80 &       3.25 & 70.8 \\ 
 &                         2007 Nov 1 06:39:10.1 &      304 &      30.2 &      30.0 &     619.28 &       3.19 & 1298.8 \\ 
 &                         2008 Nov 3 01:33:26.0 &      320 &      32.7 &      20.6 &     621.64 &       3.54 & 119.3 \\ 
 &                        2011 Dec 17 22:43:10.0 &      320 &      31.3 &      21.6 &     621.14 &       3.35 & 66.0 \\ 
 \hline
4U 1636--536 &                        2001 Sep 30 14:10:38.0 &      336 &      35.3 &      31.1 &     580.08 &       3.86 & 242.2 \\ 
 &                         2001 Oct 3 17:41:54.0 &      304 &      29.2 &      26.6 &     582.29 &       3.04 & 246.8 \\ 
 &                         2002 Jan 8 14:31:39.2 &      320 &      30.4 &      22.4 &     579.86 &       3.21 & 201.7 \\ 
 &                         2002 Jan 8 18:13:14.4 &      336 &      35.8 &      39.5 &     580.41 &       3.93 & 186.1 \\ 
 &                        2002 Jan 14 08:40:11.7 &      304 &      28.8 &      30.2 &     582.35 &       2.98 & 158.9 \\ 
 &                        2002 Feb 28 18:30:51.8 &      320 &      27.7 &      23.2 &     582.29 &       2.81 & 534.1 \\ 
 &                        2002 Mar 19 17:09:15.9 &      304 &      29.3 &      20.1 &     579.61 &       3.05 & 387.5 \\ 
 &                        2005 Aug 29 18:49:32.2 &      320 &      35.3 &      27.8 &     580.05 &       3.87 & 112.3 \\ 
 &                        2005 Aug 30 11:57:00.2 &      320 &      32.0 &      22.5 &     580.11 &       3.44 & 119.1 \\ 
 &                        \bf{2006 Apr 23 11:21:38.8\textsuperscript{e}} &      \bf{336} &      \bf{30.2} &      \bf{40.3} &     \bf{579.66} &       \bf{3.20} & \bf{188.3} \\ 
 &                        2006 Sep 12 07:42:51.0 &      304 &      28.1 &      26.2 &     580.12 &       2.87 & 99.3 \\ 
 &                        2007 Jun 20 03:08:59.2 &      320 &      33.7 &      35.3 &     582.26 &       3.67 & 246.4 \\ 
 &                        2007 Sep 28 21:15:40.4 &      320 &      33.5 &      18.6 &     580.06 &       3.64 & 184.0 \\ 
 &                        2008 Mar 15 17:32:12.9 &      336 &      30.2 &      31.5 &     580.66 &       3.19 & 273.4 \\ 
 &                         2008 Jul 3 19:28:59.0 &      304 &      27.2 &      25.4 &     581.33 &       2.73 & 225.8 \\ 
 &                        2008 Sep 11 03:57:02.1 &      304 &      30.6 &      24.7 &     581.77 &       3.25 & 145.7 \\ 
 &                        2008 Sep 25 01:50:51.0 &      304 &      27.1 &      18.7 &     580.67 &       2.71 & 434.3 \\ 
 &                        2009 Jan 17 04:31:21.9 &      304 &      31.1 &      22.1 &     579.60 &       3.31 & 127.7 \\ 
 &                        2009 Sep 10 07:56:26.0 &      336 &      31.1 &      30.1 &     580.55 &       3.31 & 126.5 \\ 
 \hline
MXB 1658--298 &                        1999 Apr 29 17:43:55.9 &      304 &      28.6 &      21.6 &     566.32 &       2.95 & 95.3 \\ 
 &                        2001 Aug 10 10:41:32.0 &      304 &      40.7 &      27.1 &     567.89 &       4.48 & 44.1 \\ 
 \hline
4U 1702--429 &                        2004 Jan 18 07:20:12.0 &      304 &      29.0 &      23.3 &     329.38 &       3.02 & 115.6 \\ 
 &                        2004 Apr 12 02:53:15.9 &      304 &      30.4 &      27.9 &     330.86 &       3.21 & 131.9 \\ 
 &                        2004 Apr 14 16:53:47.7 &      304 &      32.4 &      32.8 &     327.82 &       3.49 & 181.1 \\ 
 \hline
4U 1728--34 &                        1997 Sep 24 12:59:09.1 &      336 &      28.8 &      28.5 &     364.31 &       2.99 & 477.3 \\ 
 &                        1999 Feb 28 02:41:15.9 &      304 &      31.7 &      31.2 &     363.27 &       3.40 & 212.7 \\ 
 &                        1999 Aug 19 18:22:52.0 &      320 &      33.3 &      31.0 &     361.54 &       3.61 & 622.3 \\ 
 &                         2000 Mar 7 17:06:24.8 &      320 &      30.8 &      27.7 &     363.05 &       3.28 & 274.2 \\ 
 &                         2001 Feb 2 19:33:54.9 &      304 &      32.0 &      31.4 &     363.78 &       3.45 & 176.4 \\ 
 &                         2001 Feb 8 22:24:59.7 &      320 &      35.1 &      30.0 &     361.64 &       3.84 & 135.7 \\ 
 &                        2001 Nov 15 12:45:25.0 &      320 &      32.6 &      29.2 &     362.89 &       3.53 & 152.6 \\ 
 &                         2002 Mar 5 12:40:36.8 &      352 &      36.9 &      33.8 &     361.86 &       4.05 & 239.1 \\ 
 &                         2006 Aug 1 16:43:07.0 &      304 &      29.6 &      23.8 &     363.99 &       3.10 & 317.1 \\ 
 &                        2007 Mar 16 02:26:34.9 &      320 &      30.5 &      24.2 &     362.37 &       3.23 & 357.3 \\ 
 &                        2007 Apr 16 20:43:46.9 &      352 &      32.8 &      28.8 &     361.11 &       3.55 & 378.5 \\ 
 \hline
XTE 1739--285 &                         2005 Nov 1 01:17:25.0 &      336 &      31.0 &      24.3 &    1120.59 &       3.30 & 116.5 \\ 
 &                        2005 Nov 15 06:50:49.0 &      304 &      28.6 &      17.8 &    1123.96 &       2.95 & 85.5 \\ 
 \hline
SAX J1750.8--2900 &                        2008 May 12 07:21:28.9 &      320 &      31.3 &      24.6 &     599.44 &       3.35 & 445.9 \\ 
\hline
GS 1826--238 &                        1998 Jun 23 19:45:00.9 &      352 &      34.9 &      21.5 &     609.09 &       3.81 & 99.2 \\ 
 &                        2004 Jul 20 20:13:02.0 &      304 &      30.3 &      15.9 &     611.85 &       3.21 & 142.7 \\ 
 &                        2006 Aug 10 05:10:35.1 &      304 &      31.9 &      29.2 &     612.91 &       3.43 & 120.4 \\ 
 &                        2006 Aug 15 16:48:43.0 &      304 &      30.9 &      31.1 &     610.16 &       3.29 & 148.9 \\ 
 \hline
Aql X--1 &                        1997 Feb 27 07:10:21.9 &      304 &      31.4 &      20.6 &     549.83 &       3.36 & 204.8 \\ 
 &                        1997 Aug 28 15:46:53.0 &      320 &      39.3 &      31.9 &     548.24 &       4.33 & 439.5 \\ 
 &                        1998 Mar 10 22:48:18.9\textsuperscript{f} &      336 &      76.3 &      94.7 &     550.27 &       7.39 & 894.3 \\ 
 &                        2002 Mar 27 15:35:45.9 &      336 &      33.0 &      27.6 &     549.50 &       3.57 & 29.9 \\ 
 &                         2005 Apr 6 07:31:47.8 &      304 &      28.6 &      29.8 &     551.62 &       2.94 & 138.8 \\ 
 &                        2005 May 12 05:50:18.9 &      320 &      27.8 &      22.4 &     551.93 &       2.82 & 34.3 \\ 
 &                        2005 May 15 20:08:27.9 &      304 &      31.7 &      28.5 &     551.44 &       3.41 & 22.6 \\ 
 &                        2007 May 29 16:26:57.9 &      320 &      32.4 &      28.2 &     551.53 &       3.50 & 113.5 \\ 
 &                        2007 May 31 00:11:54.9 &      336 &      36.5 &      24.6 &     550.70 &       4.01 & 132.3 \\ 
 &                         2007 Jul 2 12:28:41.0 &      304 &      33.5 &      25.3 &     551.00 &       3.64 & 17.7 \\
\hline
\end{longtable}
\begin{flushleft} \footnotesize 
$^a$ Time (UTC) is the starting time of the candidate pulsation episode. \\
$^b$ ${t_{dur}}$ is measured from the starting time of the first time segment to the ending time of the last time segment that fits our detection criterion of 2.5${\sigma}$ statistical significance in four consecutive time segments. \\
$^c$ Leahy power, ${Z^2}$ power, ${f_{L}}$ and ${sig_{L}}$ are obtained from the time segment where the strongest pulsation is observed within the corresponding pulsation episode. \\
$^d$ Count rates during the candidate pulsation episodes in the energy range of $\sim$3-27 keV.\\
$^e$ This is the pulsation that is shown in Fig~\ref{fig X13}. \\
$^f$ This is already reported as an intermittent pulsation by \citet{Casella2008}. \\
\end{flushleft}
\twocolumn

\bibliographystyle{pasa-mnras}
\bibliography{ms}

\begin{thebibliography}{}
\makeatletter
\relax
\def\mn@urlcharsother{\let\do\@makeother \do\$\do\&\do\#\do\^\do\_\do\%\do\~}
\definecolor{darkblue}{rgb}{0,0,0.597656}
\def\mndoi{\begingroup\mn@urlcharsother \@ifnextchar [ {\mndoi@} {\mndoi@[]}}
\def\mndoi@[#1]#2{\def\@tempa{#1}\ifx\@tempa\@empty \href
  {http://dx.doi.org/#2} {\textcolor{darkblue}{doi:#2}}\else \href
  {http://dx.doi.org/#2} {\textcolor{darkblue}{#1}}\fi \endgroup}
\def\mn@eprint#1#2{\mn@eprint@#1:#2::\@nil}
\def\mn@eprint@arXiv#1{\href {http://arxiv.org/abs/#1} {{\tt arXiv:#1}}}
\def\mn@eprint@dblp#1{\href {http://dblp.uni-trier.de/rec/bibtex/#1.xml}
  {dblp:#1}}
\def\mn@eprint@#1:#2:#3:#4\@nil{\def\@tempa {#1}\def\@tempb {#2}\def\@tempc
  {#3}\ifx \@tempc \@empty \let \@tempc \@tempb \let \@tempb \@tempa \fi \ifx
  \@tempb \@empty \def\@tempb {arXiv}\fi \@ifundefined
  {mn@eprint@\@tempb}{\@tempb:\@tempc}{\expandafter \expandafter \csname
  mn@eprint@\@tempb\endcsname \expandafter{\@tempc}}}

\bibitem[\protect\citeauthoryear{{Altamirano}, {Casella}, {Patruno}, {Wijnands}
   \& {van der Klis}}{{Altamirano} et~al.}{2008}]{Altamirano2008}
{Altamirano} D.,  {Casella} P.,  {Patruno} A.,  {Wijnands} R.,   {van der Klis}
  M.,  2008, \mndoi [\apjl] {10.1086/528983}, \href
  {https://ui.adsabs.harvard.edu/abs/2008ApJ...674L..45A} {674, L45}

\bibitem[\protect\citeauthoryear{{Altamirano} et~al.,}{{Altamirano}
  et~al.}{2010}]{Altamirano2010}
{Altamirano} D.,  et~al., 2010, \mndoi [\mnras]
  {10.1111/j.1365-2966.2009.15627.x}, \href
  {https://ui.adsabs.harvard.edu/abs/2010MNRAS.401..223A} {401, 223}

\bibitem[\protect\citeauthoryear{{Andersen} \& {Ransom}}{{Andersen} \&
  {Ransom}}{2018}]{Andersen2018}
{Andersen} B.~C.,  {Ransom} S.~M.,  2018, \mndoi [\apjl]
  {10.3847/2041-8213/aad59f}, \href
  {https://ui.adsabs.harvard.edu/abs/2018ApJ...863L..13A} {863, L13}

\bibitem[\protect\citeauthoryear{{Anderson}, {Gorham}, {Kulkarni}, {Prince}  \&
  {Wolszczan}}{{Anderson} et~al.}{1990}]{Anderson1990}
{Anderson} S.~B.,  {Gorham} P.~W.,  {Kulkarni} S.~R.,  {Prince} T.~A.,
  {Wolszczan} A.,  1990, \mndoi [\nat] {10.1038/346042a0}, \href
  {https://ui.adsabs.harvard.edu/abs/1990Natur.346...42A} {346, 42}

\bibitem[\protect\citeauthoryear{{Balman}}{{Balman}}{2009}]{Balman2009}
{Balman} S.,  2009, The Astronomer's Telegram, \href
  {https://ui.adsabs.harvard.edu/abs/2009ATel.2097....1B} {2097, 1}

\bibitem[\protect\citeauthoryear{{Bhattacharyya}, {Strohmayer}, {Markwardt}  \&
  {Swank}}{{Bhattacharyya} et~al.}{2006}]{Bhattacharyya2006}
{Bhattacharyya} S.,  {Strohmayer} T.~E.,  {Markwardt} C.~B.,   {Swank} J.~H.,
  2006, \mndoi [\apjl] {10.1086/501438}, \href
  {https://ui.adsabs.harvard.edu/abs/2006ApJ...639L..31B} {639, L31}

\bibitem[\protect\citeauthoryear{{Bilous} et~al.,}{{Bilous}
  et~al.}{2019}]{Bilous2019}
{Bilous} A.~V.,  et~al., 2019, \mndoi [\apjl] {10.3847/2041-8213/ab53e7}, \href
  {https://ui.adsabs.harvard.edu/abs/2019ApJ...887L..23B} {887, L23}

\bibitem[\protect\citeauthoryear{{Blandford} \& {Teukolsky}}{{Blandford} \&
  {Teukolsky}}{1976}]{Blandford1976}
{Blandford} R.,  {Teukolsky} S.~A.,  1976, \mndoi [\apj] {10.1086/154315},
  \href {https://ui.adsabs.harvard.edu/abs/1976ApJ...205..580B} {205, 580}

\bibitem[\protect\citeauthoryear{{Brainerd} \& {Lamb}}{{Brainerd} \&
  {Lamb}}{1987}]{bra87}
{Brainerd} J.,  {Lamb} F.~K.,  1987, \mndoi [\apjl] {10.1086/184908}, \href
  {https://ui.adsabs.harvard.edu/abs/1987ApJ...317L..33B} {317, L33}

\bibitem[\protect\citeauthoryear{{Buccheri} et~al.,}{{Buccheri}
  et~al.}{1983}]{Buccheri1983}
{Buccheri} R.,  et~al., 1983, \aap, \href
  {https://ui.adsabs.harvard.edu/abs/1983A&A...128..245B} {128, 245}

\bibitem[\protect\citeauthoryear{{Casella}, {Altamirano}, {Patruno}, {Wijnands}
   \& {van der Klis}}{{Casella} et~al.}{2008}]{Casella2008}
{Casella} P.,  {Altamirano} D.,  {Patruno} A.,  {Wijnands} R.,   {van der Klis}
  M.,  2008, \mndoi [\apjl] {10.1086/528982}, \href
  {https://ui.adsabs.harvard.edu/abs/2008ApJ...674L..41C} {674, L41}

\bibitem[\protect\citeauthoryear{{Chevalier} \& {Ilovaisky}}{{Chevalier} \&
  {Ilovaisky}}{1991}]{Chevalier1991}
{Chevalier} C.,  {Ilovaisky} S.~A.,  1991, \aap, \href
  {https://ui.adsabs.harvard.edu/abs/1991A&A...251L..11C} {251, L11}

\bibitem[\protect\citeauthoryear{{Galloway}, {Muno}, {Hartman}, {Psaltis}  \&
  {Chakrabarty}}{{Galloway} et~al.}{2008}]{Galloway2008}
{Galloway} D.~K.,  {Muno} M.~P.,  {Hartman} J.~M.,  {Psaltis} D.,
  {Chakrabarty} D.,  2008, \mndoi [\apjs] {10.1086/592044}, \href
  {https://ui.adsabs.harvard.edu/abs/2008ApJS..179..360G} {179, 360}

\bibitem[\protect\citeauthoryear{{Galloway}, {Lin}, {Chakrabarty}  \&
  {Hartman}}{{Galloway} et~al.}{2010}]{Galloway2010}
{Galloway} D.~K.,  {Lin} J.,  {Chakrabarty} D.,   {Hartman} J.~M.,  2010,
  \mndoi [\apjl] {10.1088/2041-8205/711/2/L148}, \href
  {https://ui.adsabs.harvard.edu/abs/2010ApJ...711L.148G} {711, L148}

\bibitem[\protect\citeauthoryear{{G{\"o}{\v{g}}{\"u}{\textcommabelow s}},
  {Alpar}  \& {Gilfanov}}{{G{\"o}{\v{g}}{\"u}{\textcommabelow s}}
  et~al.}{2007}]{Gogus2007}
{G{\"o}{\v{g}}{\"u}{\textcommabelow s}} E.,  {Alpar} M.~A.,   {Gilfanov} M.,
  2007, \mndoi [\apj] {10.1086/512028}, \href
  {https://ui.adsabs.harvard.edu/abs/2007ApJ...659..580G} {659, 580}

\bibitem[\protect\citeauthoryear{{Hartman}, {Chakrabarty}, {Galloway}, {Muno},
  {Savov}, {Mendez}, {van Straaten}  \& {Di Salvo}}{{Hartman}
  et~al.}{2003}]{Hartman2003}
{Hartman} J.~M.,  {Chakrabarty} D.,  {Galloway} D.~K.,  {Muno} M.~P.,  {Savov}
  P.,  {Mendez} M.,  {van Straaten} S.,   {Di Salvo} T.,  2003, in AAS/High
  Energy Astrophysics Division \#7. AAS/High Energy Astrophysics Division.
p. 17.38

\bibitem[\protect\citeauthoryear{{Jain} \& {Paul}}{{Jain} \&
  {Paul}}{2011}]{Jain2011}
{Jain} C.,  {Paul} B.,  2011, \mndoi [Research in Astronomy and Astrophysics]
  {10.1088/1674-4527/11/5/007}, \href
  {https://ui.adsabs.harvard.edu/abs/2011RAA....11..577J} {11, 577}

\bibitem[\protect\citeauthoryear{{Kaaret}, {in 't Zand}, {Heise}  \&
  {Tomsick}}{{Kaaret} et~al.}{2002}]{Kaaret2002}
{Kaaret} P.,  {in 't Zand} J.~J.~M.,  {Heise} J.,   {Tomsick} J.~A.,  2002,
  \mndoi [\apj] {10.1086/341336}, \href
  {https://ui.adsabs.harvard.edu/abs/2002ApJ...575.1018K} {575, 1018}

\bibitem[\protect\citeauthoryear{{Kaaret} et~al.,}{{Kaaret}
  et~al.}{2007}]{Kaaret2007}
{Kaaret} P.,  et~al., 2007, \mndoi [\apjl] {10.1086/513270}, \href
  {https://ui.adsabs.harvard.edu/abs/2007ApJ...657L..97K} {657, L97}

\bibitem[\protect\citeauthoryear{{Kulkarni} \& {Romanova}}{{Kulkarni} \&
  {Romanova}}{2008}]{kul08}
{Kulkarni} A.~K.,  {Romanova} M.~M.,  2008, \mndoi [\mnras]
  {10.1111/j.1365-2966.2008.13094.x}, \href
  {https://ui.adsabs.harvard.edu/abs/2008MNRAS.386..673K} {386, 673}

\bibitem[\protect\citeauthoryear{{Leahy}, {Darbro}, {Elsner}, {Weisskopf},
  {Sutherland}, {Kahn}  \& {Grindlay}}{{Leahy} et~al.}{1983}]{Leahy1983}
{Leahy} D.~A.,  {Darbro} W.,  {Elsner} R.~F.,  {Weisskopf} M.~C.,  {Sutherland}
  P.~G.,  {Kahn} S.,   {Grindlay} J.~E.,  1983, \mndoi [\apj] {10.1086/160766},
  \href {https://ui.adsabs.harvard.edu/abs/1983ApJ...266..160L} {266, 160}

\bibitem[\protect\citeauthoryear{{Markwardt}, {Strohmayer}  \&
  {Swank}}{{Markwardt} et~al.}{1999}]{Markwardt1999}
{Markwardt} C.~B.,  {Strohmayer} T.~E.,   {Swank} J.~H.,  1999, \mndoi [\apjl]
  {10.1086/311886}, \href
  {https://ui.adsabs.harvard.edu/abs/1999ApJ...512L.125M} {512, L125}

\bibitem[\protect\citeauthoryear{{Messenger}}{{Messenger}}{2011}]{Messenger2011}
{Messenger} C.,  2011, \mndoi [\prd] {10.1103/PhysRevD.84.083003}, \href
  {https://ui.adsabs.harvard.edu/abs/2011PhRvD..84h3003M} {84, 083003}

\bibitem[\protect\citeauthoryear{{Messenger} \& {Patruno}}{{Messenger} \&
  {Patruno}}{2015}]{Messenger2015}
{Messenger} C.,  {Patruno} A.,  2015, \mndoi [\apj]
  {10.1088/0004-637X/806/2/261}, \href
  {https://ui.adsabs.harvard.edu/abs/2015ApJ...806..261M} {806, 261}

\bibitem[\protect\citeauthoryear{{Middleditch} \& {Kristian}}{{Middleditch} \&
  {Kristian}}{1984}]{Middleditch1984}
{Middleditch} J.,  {Kristian} J.,  1984, \mndoi [\apj] {10.1086/161876}, \href
  {https://ui.adsabs.harvard.edu/abs/1984ApJ...279..157M} {279, 157}

\bibitem[\protect\citeauthoryear{{Muno}, {Fox}, {Morgan}  \& {Bildsten}}{{Muno}
  et~al.}{2000}]{Muno2000}
{Muno} M.~P.,  {Fox} D.~W.,  {Morgan} E.~H.,   {Bildsten} L.,  2000, \mndoi
  [\apj] {10.1086/317031}, \href
  {https://ui.adsabs.harvard.edu/abs/2000ApJ...542.1016M} {542, 1016}

\bibitem[\protect\citeauthoryear{{{\"O}zel}}{{{\"O}zel}}{2009}]{Ozel2009}
{{\"O}zel} F.,  2009, \mndoi [\apj] {10.1088/0004-637X/691/2/1678}, \href
  {https://ui.adsabs.harvard.edu/abs/2009ApJ...691.1678O} {691, 1678}

\bibitem[\protect\citeauthoryear{{Patruno} \& {Watts}}{{Patruno} \&
  {Watts}}{2012}]{PatrunoWatts2012}
{Patruno} A.,  {Watts} A.~L.,  2012, arXiv e-prints, \href
  {https://ui.adsabs.harvard.edu/abs/2012arXiv1206.2727P} {p. arXiv:1206.2727}

\bibitem[\protect\citeauthoryear{{Patruno}, {Wette}  \& {Messenger}}{{Patruno}
  et~al.}{2018}]{Patruno2018}
{Patruno} A.,  {Wette} K.,   {Messenger} C.,  2018, \mndoi [\apj]
  {10.3847/1538-4357/aabf89}, \href
  {https://ui.adsabs.harvard.edu/abs/2018ApJ...859..112P} {859, 112}

\bibitem[\protect\citeauthoryear{{Ransom}, {Greenhill}, {Herrnstein},
  {Manchester}, {Camilo}, {Eikenberry}  \& {Lyne}}{{Ransom}
  et~al.}{2001}]{Ransom2001}
{Ransom} S.~M.,  {Greenhill} L.~J.,  {Herrnstein} J.~R.,  {Manchester} R.~N.,
  {Camilo} F.~o.,  {Eikenberry} S.~S.,   {Lyne} A.~G.,  2001, \mndoi [\apjl]
  {10.1086/318062}, \href
  {https://ui.adsabs.harvard.edu/abs/2001ApJ...546L..25R} {546, L25}

\bibitem[\protect\citeauthoryear{{Ransom}, {Eikenberry}  \&
  {Middleditch}}{{Ransom} et~al.}{2002}]{Ransom2002}
{Ransom} S.~M.,  {Eikenberry} S.~S.,   {Middleditch} J.,  2002, \mndoi [\aj]
  {10.1086/342285}, \href
  {https://ui.adsabs.harvard.edu/abs/2002AJ....124.1788R} {124, 1788}

\bibitem[\protect\citeauthoryear{{Smith}, {Morgan}  \& {Bradt}}{{Smith}
  et~al.}{1997}]{Smith1997}
{Smith} D.~A.,  {Morgan} E.~H.,   {Bradt} H.,  1997, \mndoi [\apjl]
  {10.1086/310604}, \href
  {https://ui.adsabs.harvard.edu/abs/1997ApJ...479L.137S} {479, L137}

\bibitem[\protect\citeauthoryear{{Strohmayer} \& {Markwardt}}{{Strohmayer} \&
  {Markwardt}}{2002}]{Strohmayer2002}
{Strohmayer} T.~E.,  {Markwardt} C.~B.,  2002, \mndoi [\apj] {10.1086/342152},
  \href {https://ui.adsabs.harvard.edu/abs/2002ApJ...577..337S} {577, 337}

\bibitem[\protect\citeauthoryear{{Strohmayer}, {Zhang}, {Smale}, {Day},
  {Swank}, {Titarchuk}  \& {Lee}}{{Strohmayer} et~al.}{1996}]{Strohmayer1996}
{Strohmayer} T.,  {Zhang} W.,  {Smale} A.,  {Day} C.,  {Swank} J.,  {Titarchuk}
  L.,   {Lee} U.,  1996, \iaucirc, \href
  {https://ui.adsabs.harvard.edu/abs/1996IAUC.6387....2S} {6387, 2}

\bibitem[\protect\citeauthoryear{{Strohmayer}, {Zhang}, {Swank}, {White}  \&
  {Lapidus}}{{Strohmayer} et~al.}{1998}]{Strohmayer1998}
{Strohmayer} T.~E.,  {Zhang} W.,  {Swank} J.~H.,  {White} N.~E.,   {Lapidus}
  I.,  1998, \mndoi [\apjl] {10.1086/311322}, \href
  {https://ui.adsabs.harvard.edu/abs/1998ApJ...498L.135S} {498, L135}

\bibitem[\protect\citeauthoryear{{Strohmayer} et~al.,}{{Strohmayer}
  et~al.}{2018}]{Strohmayer2018}
{Strohmayer} T.~E.,  et~al., 2018, \mndoi [\apjl] {10.3847/2041-8213/aabf44},
  \href {https://ui.adsabs.harvard.edu/abs/2018ApJ...858L..13S} {858, L13}

\bibitem[\protect\citeauthoryear{{Taylor} \& {Weisberg}}{{Taylor} \&
  {Weisberg}}{1989}]{Taylor1989}
{Taylor} J.~H.,  {Weisberg} J.~M.,  1989, \mndoi [\apj] {10.1086/167917}, \href
  {https://ui.adsabs.harvard.edu/abs/1989ApJ...345..434T} {345, 434}

\bibitem[\protect\citeauthoryear{{Thompson}, {Rothschild}, {Tomsick}  \&
  {Marshall}}{{Thompson} et~al.}{2005}]{Thompson2005}
{Thompson} T. W.~J.,  {Rothschild} R.~E.,  {Tomsick} J.~A.,   {Marshall} H.~L.,
   2005, \mndoi [\apj] {10.1086/497104}, \href
  {https://ui.adsabs.harvard.edu/abs/2005ApJ...634.1261T} {634, 1261}

\bibitem[\protect\citeauthoryear{{Titarchuk}, {Cui}  \& {Wood}}{{Titarchuk}
  et~al.}{2002}]{tit02}
{Titarchuk} L.,  {Cui} W.,   {Wood} K.,  2002, \mndoi [\apjl] {10.1086/343099},
  \href {https://ui.adsabs.harvard.edu/abs/2002ApJ...576L..49T} {576, L49}

\bibitem[\protect\citeauthoryear{{Villarreal} \& {Strohmayer}}{{Villarreal} \&
  {Strohmayer}}{2004}]{Villarreal2004}
{Villarreal} A.~R.,  {Strohmayer} T.~E.,  2004, \mndoi [\apjl]
  {10.1086/425737}, \href
  {https://ui.adsabs.harvard.edu/abs/2004ApJ...614L.121V} {614, L121}

\bibitem[\protect\citeauthoryear{{Wachter}, {Hoard}, {Bailyn}, {Corbel}  \&
  {Kaaret}}{{Wachter} et~al.}{2002}]{Wachter2002}
{Wachter} S.,  {Hoard} D.~W.,  {Bailyn} C.~D.,  {Corbel} S.,   {Kaaret} P.,
  2002, \mndoi [\apj] {10.1086/339034}, \href
  {https://ui.adsabs.harvard.edu/abs/2002ApJ...568..901W} {568, 901}

\bibitem[\protect\citeauthoryear{{Watts}}{{Watts}}{2012}]{Watts2012}
{Watts} A.~L.,  2012, \mndoi [\araa] {10.1146/annurev-astro-040312-132617},
  \href {https://ui.adsabs.harvard.edu/abs/2012ARA&A..50..609W} {50, 609}

\bibitem[\protect\citeauthoryear{{Welsh}, {Robinson}  \& {Young}}{{Welsh}
  et~al.}{2000}]{Welsh2000}
{Welsh} W.~F.,  {Robinson} E.~L.,   {Young} P.,  2000, \mndoi [\aj]
  {10.1086/301486}, \href
  {https://ui.adsabs.harvard.edu/abs/2000AJ....120..943W} {120, 943}

\bibitem[\protect\citeauthoryear{{Wijnands}, {Strohmayer}  \&
  {Franco}}{{Wijnands} et~al.}{2001}]{Wijnands2001}
{Wijnands} R.,  {Strohmayer} T.,   {Franco} L.~M.,  2001, \mndoi [\apjl]
  {10.1086/319128}, \href
  {https://ui.adsabs.harvard.edu/abs/2001ApJ...549L..71W} {549, L71}

\bibitem[\protect\citeauthoryear{{Wood}, {Ftaclas}  \& {Kearney}}{{Wood}
  et~al.}{1988}]{Wood1988}
{Wood} K.~S.,  {Ftaclas} C.,   {Kearney} M.,  1988, \mndoi [\apjl]
  {10.1086/185092}, \href
  {https://ui.adsabs.harvard.edu/abs/1988ApJ...324L..63W} {324, L63}

\bibitem[\protect\citeauthoryear{{Wood} et~al.,}{{Wood}
  et~al.}{1991}]{Wood1991}
{Wood} K.~S.,  et~al., 1991, \mndoi [\apj] {10.1086/170505}, \href
  {https://ui.adsabs.harvard.edu/abs/1991ApJ...379..295W} {379, 295}

\bibitem[\protect\citeauthoryear{{Zhang}, {Jahoda}, {Kelley}, {Strohmayer},
  {Swank}  \& {Zhang}}{{Zhang} et~al.}{1998}]{Zhang1998}
{Zhang} W.,  {Jahoda} K.,  {Kelley} R.~L.,  {Strohmayer} T.~E.,  {Swank} J.~H.,
    {Zhang} S.~N.,  1998, \mndoi [\apjl] {10.1086/311210}, \href
  {https://ui.adsabs.harvard.edu/abs/1998ApJ...495L...9Z} {495, L9}

\bibitem[\protect\citeauthoryear{{van Haaften}, {Nelemans}, {Voss}, {van der
  Sluys}  \& {Toonen}}{{van Haaften} et~al.}{2015}]{vanHaaften15}
{van Haaften} L.~M.,  {Nelemans} G.,  {Voss} R.,  {van der Sluys} M.~V.,
  {Toonen} S.,  2015, \mndoi [\aap] {10.1051/0004-6361/201425303}, \href
  {https://ui.adsabs.harvard.edu/abs/2015A&A...579A..33V} {579, A33}

\bibitem[\protect\citeauthoryear{{van der Klis}, {Wijnands}, {Horne}  \&
  {Chen}}{{van der Klis} et~al.}{1997}]{Klis1997}
{van der Klis} M.,  {Wijnands} R. A.~D.,  {Horne} K.,   {Chen} W.,  1997,
  \mndoi [\apjl] {10.1086/310656}, \href
  {https://ui.adsabs.harvard.edu/abs/1997ApJ...481L..97V} {481, L97}

\makeatother
\end{thebibliography}

\end{document}